\documentclass[twocolumn]{aastex}
\usepackage{amssymb,amsmath}
\def\kms{km s$^{-1}$}
\def\cc{${\rm cm^{-3}}$}

\def\htwo{${\rm H_2}$}
\def\nht{${\rm NH_3}$}
\def\um{\rm \mu m}

\shorttitle{Core evolution in NGC 2024}
\shortauthors{Ren et al.}

\begin{document}
\title{Massive Quiescent Cores in Orion: VI. The Internal Structures and a Candidate of Transiting Core in NGC 2024 Filament}
\author{Zhiyuan Ren\altaffilmark{1,2,3} and Di Li\altaffilmark{1,2}}

\altaffiltext{1}{National Astronomical Observatories, Chinese Academy of
	Science, Chaoyang District Datun Rd A20, Beijing, China; Email:
	renzy@nao.cas.cn, dili@nao.cas.cn}
\altaffiltext{2}{Key Laboratory of Radio Astronomy, Chinese Academy of Science}
\altaffiltext{3}{The Department of Astronomy, Peking University}
 

\begin{abstract}
We present a multi-wavelength observational study of the NGC 2024 filament using infrared to sub-millimeter continuum and the \nht\ $(1,1)$ and $(2,2)$ inversion transitions centered on FIR-3, the most massive core therein. FIR-3 is found to have no significant infrared point sources in the Spitzer/IRAC bands. But the \nht\ kinetic temperature map shows a peak value at the core center with $T_{\rm k}=25$ K which is significantly higher than the surrounding level ($T_{\rm k}=15-19$ K). Such internal heating signature without an infrared source suggests an ongoing core collapse possibly at a transition stage from first hydrostatic core (FHSC) to protostar. The eight dense cores in the filament have dust temperatures between 17.5 and 22 K. They are much cooler than the hot ridge ($T_{\rm d}=55$ K) around the central heating star IRS-2b. Comparison with a dust heating model suggests that the filament should have a distance of $3-5$ pc from IRS-2b. This value is much larger than the spatial extent of the hot ridge, suggesting that the filament is spatially separated from the hot region along the line of sight.
\end{abstract}


\section{Introduction}   
Revealing the transition stage between prestellar cores and protostars is critical for understanding the entire star forming process. The low- and intermediate-mass prestellar cores are supposed to initially stay in a hydrostatic equilibrium, referred to as Bonnor-Ebert sphere \citep{ebert55,bonnor56}. The density would gradually increase due to the self contraction and external influences such as turbulence, compression flow, and velocity perturbation which are usually presented in filamentary clouds \citep[e.g.][]{gomez07,hennebelle11,gong15}. As the core becomes supercritical to the self gravity, the "first collapse" would occur and generate a much denser object called the first hydrostatic core (FHSC, or first core). The FHSC has a density of $\sim10^{13}$ \cc\ so that the gas and dust become opaque to radiation \citep{larson69}. The temperature in the FHSC would continuously increase due to the self-gravity and possibly on-going accretion. Once the central temperature exceeds 2000 K, the heating would start to dissociate the \htwo\ molecules. The \htwo\ dissociation can provide an efficient coolant for the gas, and largely reduce the thermal pressure support, thereby induce the "second collapse". In this process the FHSC would evolve into a protostar \citep{masunaga00,andre07}. 


Despite this delicately modelled evolutionary track, the observed examples for the transition stage between pre- and protostellar cores are still scarce. Up till now, only a few low-mass cores are suggested to be the FHSC candidates \citep[e.g.][]{boss95,belloche06,chen10,pineda11,pezzuto12}. And a few properties are expected for the FHSCs based on their dust continuum emissions, including: (1) low luminosity ($<10^{-1}~L_\odot$) and temperature ($<20$ K); (2) no significant emissions in far-infrared and shorter wavelengths ($\lambda<70~\um$); (3) dense and compact morphology in (sub)millimeter wavelengths.

At a distance of 415 pc \citep{menten07,sandstrom07}, Orion molecular cloud is the closest and best-studied massive star-forming complex. Besides the regions with bright young stars, Orion contains a huge amount of cold, quiescent, and dense gas \citep[e.g.][and references therein]{salji15}. In our previous studies \citep{li07,velusamy08,li13,ren14}, the quiescent cores in Orion A and Orion South have been selected and investigated. A large fraction of the cores were found to be unstable to the self-gravity, and likely to have lower temperatures than their surroundings \citep{li13}. The candidate cores for the transition stage can be selected based on two requirements: the cores should be supercritical to the self-gravity and meanwhile have no detectable IR sources. Some likely candidates were examined but found to actually have faint embedded IR sources and even the multiple stars which are associated with the core fragmentation \citep{ren14}. Subsequent studies should be performed over a larger field in order to enlarge the sample. And the evolutionary stages should be evaluated based on more evidences. 

Located $\sim4$ degrees to the North of Orion A, NGC 2024 in Orion B cloud contains extended gas structures with embedded cold dense cores. The major fraction of the gas is assembled in a compact filamentary structure \citep[][also see Figure 1]{mezger92}. A number of observations were performed to examine the physical properties therein \citep{gaume92,mezger92,mauersberger92,chandler96,watanabe08,alves11,choi15}. These observations revealed that the dense filament is located behind the hot ionized gas and have complex structures. But the specific properties and evolutionary state of each core are still to be investigated. \citet{gaume92} presented Very Large Array (VLA) observation in the \nht\ lines, but only obtained average physical parameters for several regions owing to the limited velocity resolution and spectral sensitivity. 

In this work, we presented new observational study for NGC 2024 using the continuum emissions from mid-infrared to sub-millimeter bands. We examined the mass and the temperature distributions of the entire filament. Moreover, to study the star-forming properties in the most massive core FIR-3 therein, we carried out new \nht\ observation with the Karl G. Jansky Very Large Array (JVLA). The results show that the dense cores on the filament tend to have young evolutionary stages and should be spatially separated from the hot gas and dust in the foreground. In Section 2, we described the observation. In Section 3 and 4, we described the filament and dense core structures and their dust temperatures. In Section 5, we presented the mass and temperature distributions within FIR-3 based on the high-resolution \nht\ spectral data. In Section 6, we discussed the core evolutionary state based on their physical parameters. A summary is given in Section 7.

\section{Observation and Data Reduction} 
The observation of NGC 2024 FIR-3 was carried out with the NRAO\footnote{\scriptsize The
National Radio Astronomy Observatory is a facility of the National Science
Foundation operated under cooperative agreement by Associated Universities, Inc.}
JVLA on September 6, 2014. The antennae were in D configuration. One baseband (A0/C0) was tuned at 23.9 GHz, in which two sub-bands were placed at the frequencies of the \nht\ $(1,1)$ and $(2,2)$ lines, respectively. Each subband had a bandwidth of 8 MHz. All the hyperfine components (HFCs) of the \nht $(1,1)$ lines are covered by the subband with a velocity resolution of 0.1 \kms. 3C147, which has a flux density of 2.8 Jy in the K-band under the D-configuration, was used as the bandpass and flux calibrator. The quasar J0541-0541 was adopted as the antenna gain calibrator. The total on-source integration time was 50 minutes. The CASA program\footnote{\scriptsize http://casa.nrao.edu} and  were used for the data calibration analysis and imaging. Aladin sky atlas\footnote{\scriptsize "Aladin sky atlas" is developed at CDS, Strasbourg Observatory, France \citep{boch14}} was also used for inspecting the images. The synthesized beam size is $3.3''\times2.8''$ and the rms noise level is 7 mJy beam$^{-1}$ (1.2 K) per 0.1 \kms\ channel.

In order to investigate the entire filament structure, the stellar emissions, and the spectral energy distribution (SED), we also obtained mid- to far-infrared images including (1) Spitzer/IRAC images from the Archive of the Spitzer Enhanced Imaging Products (SEIP) \footnote{\scriptsize http://sha.ipac.caltech.edu/applications/Spitzer/SHA/}, (2) The Herschel PACS 70, 100, and 160 $\um$, SPIRE 250, 350, and 500 $\um$ images taken from the Herschel Science Archive\footnote{\scriptsize Herschel is an ESA space observatory with science instruments provided by European-led Principal Investigator consortia and with important participation from NASA. http://www.cosmos.esa.int/web/herschel/science-archive}. The Herschel observations are part of the Gould Belt Survey \citep{andre10}. (3) The MSX 12, 16.5 and 21 $\um$ images from the MSX Image Server and Catalog Overlays v 6.0\footnote{\scriptsize http://irsa.ipac.caltech.edu/applications/MSX/MSX/}, and (4) The JCMT/SCUBA 450 and 850 $\um$ continuum maps from the JCMT science archive\footnote{\scriptsize The James Clerk Maxwell Telescope is operated by the Joint Astronomy Centre on behalf of the Science and Technology Facilities Council of the United Kingdom, the Netherlands Organisation for Scientific Research, and the National Research Council of Canada. http://www.cadc.hia.nrc.gc.ca/jcmt/}. 

The Herschel PACS and SPIRE images are Level-2.5 and Level-3 products, respectively. The PACS 100 and 160 $\um$ images are from the "PACS-only" data and were mapped at a scan speed of 20 arcsec s$^{-1}$ and Repetition factor of 6. The PACS 70 $\um$ and all the SPIRE images are in parallel mode and have a scan speed of 60 arcsec s$^{-1}$. The Herschel images have sensitivities between 5 and 40 mJy pixel$^{-1}$ (see Table A1), which are 2 to 3 orders of magnitudes smaller than the filament emissions. Therefore, the flux uncertainties should mainly come from the flux calibration and photometry processes, as described in Appendix A.1..


\section{The Mass Distribution of the Filament and the Ridge}  
\subsection{Dust Continuum Emissions} 
The Spitzer, MSX, Herschel, and JCMT/SCUBA images from mid-infrared to sub-millimeter bands are shown in Figure 1. Originally, the PACS 70, 100, and 160 $\um$ images are in unit of Jy pixel$^{-1}$, and the SPIRE and SCUBA images are in Jy beam$^{-1}$. In Figure 1, the intensity unit was scaled to Jy arcsec$^{-2}$ by dividing the original intensity scale with the pixel (PACS) or the beam (SPIRE and SCUBA) areas, in order to provide a fair comparison for the intensity scales throughout the wavelength range. 

Figure 1a shows the SCUBA 450 $\um$ emission (contours) overlaid on the RGB image of the IRAC 3.6, 4.5 and 8.0 $\um$ bands. The brightest star and major heating source in this region is IRS-2b \citep{bik03}, which was estimated to have a spectral type between O8V and B2V. It is located at a projected distance of $30''$ (0.05 pc) from the filament. The FIR cores (specified in Figure 2 and Section 3.2) are labeled with '+' symbols. The 8 $\um$ emission (red) shows the extended emission over the entire region, and a more prominent "hot ridge" structure going across FIR-5. The MSX RGB image (12, 16.5, and 21 $\um$ bands, shown in Figure 1b) also evidently traces the hot ridge. In the MIPS 24 $\um$ band (Figure 1c), the extended emission region is largely saturated over a spatial extent of 100-200 arcsec (0.2-0.4 pc), and its southern boundary is almost in parallel with the hot ridge. The extended emission and the ridge should represent the H{\sc ii} and the photo dissociation region (PDR) around IRS-2b, respectively \citep{roshi14}; the ridge might be formed as the H{\sc ii} region is expanding and compressing the surrounding medium. 

The PACS 70 $\um$ image (Figure 1d) is also dominated by the hot-ridge emission. Among the seven major dense cores identified in \citet{mezger92}, only FIR-4 is significantly seen. FIR-5 is totally blended with the ridge emission and cannot be separated. The PACS 100 $\um$ image (Figure 1e) exhibits similar features but with the filament more evidently seen (in particular FIR-2 and 3), while the ridge becomes less intense. In PACS 160 $\um$, SPIRE 250 $\um$, 350 $\um$, and 500 $\um$, and SCUBA 450 $\um$ and 850 $\um$ bands (Figure 1f to 1k), the emissions are dominated by the filament and exhibits similar morphologies, while the ridge is much weaker in those bands. 

\subsection{Core Identification and Flux Measurement} 
The filament and FIR cores are best revealed in the SCUBA 450 $\um$ band. On the 450 $\micron$ image, the dense cores were identified using the IDL routine Hyper \citep{traficante15}. Hyper will extract local emission peaks, identify the cores, fit the angular size and surface brightness of each core with 2D Gaussian profile, and estimate the flux density within the core area. In this process, the background emission is subtracted through a 2D polynomial fitting. The best-fit core radii for the 450 $\um$ cores are shown in Figure 2a. Using Hyper, the seven FIR cores \citep{mezger92} were all clearly extracted. In addition, FIR-5 was resolved into two objects (denoted as FIR-5a and 5b), which are consistent with the previous observations \citep{wiesemeyer97,lai02}. FIR-5a was further resolved into at least seven condensations in \citet{lai02}. 

As shown in Table 1, FIR-2, 6, 7 are fitted to have large ratios of $r_{\rm maj}/r_{\rm min}$. Such elongated core areas should be a result of confusion with the extended filamentary structures. As shown in Figure 2a, the cores are evidently seen only above the 30\% contour, while bellow this level, the emission is mainly from the filament. Figure 2 also shows that the elongations of the cores are altogether reasonably along the filament (both at 160 and 450 $\micron$). To eliminate the confusion, we performed another fitting assuming each core to have a circular shape. The measured core radii are shown in Table 1, and the circular-shaped core areas are shown in Figure 2b. It is within our expectation that the radius of the circular fitting is closer to the minor axis of the elliptical core area. 


In Figure 2b, the 450 $\um$ emission is overlaid on the IRAC RGB image, and the IRAC point sources in the Orion protostar catalogue \citep{megeath12} are also plotted with yellow stars. Only FIR-1, 2, and 4 coincide with the IRAC sources, with the magnitudes of $M_{3.6\um}=8.69$ to 9.09, as listed in Table 2. The brightest IR source coincides with FIR-4. Based on the detection limit, the FIR cores absent of IRAC sources should have $M_{3.6\um}>14.6$. 

The source extraction with Hyper was also applied for the PACS 70 $\um$ and 160 $\um$ images (Figure 2c and 2d, respectively). The 70 $\um$ image was examined for potential additional hot-dust sources, while the 160 $\um$ image was adopted to compare with the 450-$\um$ cores. As a result, FIR-4, 6 and other two sources were identified at PACS 70 $\um$. But the two sources on the ridge should represent the emission peaks of the hot dust rather than dense cores. On the PACS 160 $\um$ image, all the dense cores were identified, with the central positions consistent with the 450-$\um$ values within $7''$ (see Table 1). 

We adopted the 450-$\um$ core areas (in the circular case) to measure the flux densities of all the cores except FIR-5a and 5b. At 70 $\micron$, the emission intensities within the FIR cores (except FIR-4 and 5) range from 1 to 3 Jy arcsec$^{-2}$ which are much higher than the detection limit, but are comparable with the extended emission level ($1.5\pm0.9$ Jy arcsec$^{-2}$). We thus suggested a marginal detection for these cores above the extended emission level. The flux measurement is described in more detail in Appendix A.1. We also selected five positions on the hot gas ridge including one at FIR-5 (Figure 3b), and measured the flux density at each position within a square region that covers the ridge width ($d=25$ arcsec). On the Spitzer and MSX images (3.6 to 21 $\um$), there are no structures likely associated with the filament. The measured flux densities within the core areas should represent upper limits for the FIR cores. The flux densities of the cores in the IRAC and MSX bands are listed in Table 2, and the values of Hershel and SCUBA bands are shown in Table 3. 

\section{Dust Temperature Distribution}  
\subsection{The SED fitting for the cores and the ridge}  
We estimated the dust temperature in the FIR cores and the ridge from their spectral energy distributions (SEDs) throughout the MSX, Herschel and JCMT bands. The SED is fitted using a grey-body emission model \citep{hildebrand83}. Based on the radiative transfer, the flux density of the dust core is
\begin{equation} 
S_{\nu}=\Omega B_{\nu}(T_{\rm d})[1-\exp(-\tau_\nu)],
\end{equation}
wherein $S_{\nu}$ is the flux density at the frequency $\nu$. $\Omega$ is the solid angle of the core or the selected area. $B_{\nu}(T_{\rm d})$ is the Planck function of the dust temperature $T_{\rm d}$. The optical depth $\tau_\nu$ is determined by other parameters as  
\begin{equation}  
\tau_\nu =\kappa_\nu \mu m_{\rm H} N_{\rm tot}/g,
\end{equation}
wherein $N_{\rm tot}$ is the gas column density (mostly HI+\htwo), $\mu=2.33$ is the mean molecular weight \citep{myers83}, $m_{\rm H}$ is the mass of the hydrogen atom, and $g=100$ is the gas-to-dust mass ratio. $\kappa_\nu$ is the dust opacity, and is expected to vary with the frequency in the form $\kappa_\nu=\kappa_{\rm 230 GHz}(\nu/{\rm 230 GHz})^{\beta}$, with the reference value of $\kappa_{\rm230 GHz}=0.9$ cm$^2$ g$^{-1}$, as adopted from dust model for the grains with coagulation for $10^5$ years with accreted ice mantles at a density of $10^6$ \cc\ \citep{ossenkopf94}. 

The best-fit SEDs for the FIR cores are shown in Figure 3a (FIR-5 is put in Figure 3b to compare with the ridge). The physical parameters ($\beta$ and $T_{\rm d}$) are presented in Table 4. The cores were measured to have a small variation both in $T_{\rm d}$ and $\beta$. FIR-4 has the highest temperature of $T_{\rm d}=22$ K, which is in agreement with the brightest infrared source therein. The total luminosities of the cores were estimated using $L_{\rm core}=\int 4\pi D^2 S_\nu d\nu$, and are also presented in Table 4. We note that FIR-3 was also observed in 1.2 mm continuum \citep{hill05} with the flux density measured to be $S_{\rm 1.2 mm}=10$ Jy. In comparison, by extrapolating the SED curve, we obtained a much lower value of $S_{\rm 1.2 mm}=6$ Jy. The difference is probably due to the low resolution ($24''$) and large mapping step ($44''$) in the previous observation. The mapping step is larger than distances of FIR-1 and 4 from FIR-3, thus the cores would be poorly resolved and emissions from FIR-1 and 4 might be largely included in the measured flux density. The $\beta$ values in the filament are comparable to the highest $\beta$ values among the other Orion regions \citep{goldsmith97}. 

The SED within FIR-5 can be well fitted using two temperature components with $T_{\rm d}=17.5$ and 55 K respectively (Figure 3b). The cold component is comparable to the other FIR cores and should represent the SED of FIR-5 itself. The hot component ($T_{\rm d}=55$ K) should represent the contribution from the hot ridge. The average flux densities of the four other positions on the ridge are also shown in Figure 3b. The SED fitting resulted in $T_{\rm d}=56$ K and $\beta\simeq1.6$. We note that the $\beta$ value for the hot component at FIR-5 is not well constrained because the Rayleigh-Jeans tail of the SED is dominated by the cold component. We expect it to have a similar value of $\beta\simeq1.6$.

The optical depth from the for the hot component is optically thin throughout the observed wavelengths, with $\tau\simeq0.1$ at $\lambda=70~\um$, while the cold dust becomes moderately optically thick at $\lambda<100~\um$. At low optical depth, Equation (1) can be approximated as  
\begin{equation} 
\begin{aligned}
S_{\nu} & = \kappa_\nu B_{\nu}(T_{\rm d}) \Omega \mu m_{\rm H} N_{\rm tot}/g \\
        & = \frac{\kappa_{\nu} B_{\nu}(T_{\rm d}) M_{\Omega} }{g D^2},
\end{aligned}
\end{equation}
wherein $D=415$ pc is the source distance, $M_{\Omega}$ is the gas mass within $\Omega$. $N_{\rm tot}$ was estimated from the 450 $\micron$ peak intensity (in Jy arcsec$^{-2}$), and the masses of FIR cores and the ridge (within the square aperture region) are calculated from their total $S_{450 \micron}$ values. We note that FIR-5a and 5b are resolved at 450 $\micron$, so their masses can be separately obtained using the average $T_{\rm d}$ and $\beta$ values for the two cores. The volume number density can be estimated from $N_{\rm tot}$ as $n=N_{\rm tot}/2R_{\rm core}$. This value represents an average along the line of sight. The results are also presented in Table 4. We note that if $\kappa_{\rm 230 GHz}$ value is different, the core mass would vary with $\kappa_{\rm 230 GHz}$ proportionally, while the temperature fit remains unchanged. For example, if adopting $\kappa_{\rm 230 GHz}=0.5$ cm$^2$ g$^{-1}$ \citep{preibisch93}, the core masses would increase by a factor of $0.9/0.5=1.8$. The SED fitting is described in more details in Appendix A.2..

For the hot component, the IRAC 8 $\um$ and MSX 12.5 $\um$ intensities are above the SED curve, suggesting a dust component with even higher temperature. \citet{mezger92} fitted the SED of the entire region and identified two cold components with $T_{\rm d}=19$ and 22 K, which are consistent with the temperature range for the FIR cores. But the hot component was measured to have $T_{\rm d}=45$ K, which is cooler than the hot ridge observed here. The difference is probably because the flux densities at shorter wavelengths ($\lambda<350~\um$) were not included in their estimation. 

It is also noteworthy that the ridge has a much lower $\beta$ than the values in the FIR cores. To examine the validity of this discrepancy, we attempted to fit the SED using a fixed value ($\beta=2.7$) comparable to the FIR cores. In this case, the Rayleigh-Jeans tail ($\lambda=200$ to 1000 $\um$) of the SED largely deviates from the observed values and cannot be reconciled by adjusting $\tau$ and $T_{\rm d}$. The difference in $\beta$ may indicate different dust properties in the two components, and support the trend that $\beta$ decreases with $T_{\rm d}$ \citep{arab12}.  

\section{The \nht\ Gas Structure and Temperature Distribution}  
\subsection{The Gas Distribution}  
The \nht\ line spectra at the center of FIR-3 are shown in Figure 4, with the hyperfine structures (HFS) labelled on the lines. The \nht\ maps around FIR-3 are shown in Figure 5. The \nht\ (1,1) inner satellite group {\bf (isg)} has lower opacities than the main group {\bf (mg)} thus would better reveal the dense gas structures. Figure 5a shows the integrated emission of the {\bf isg}, which is overlaid on the IRAC image. The \nht\ gas is located around the FIR-3 core center and aligned in parallel with the 450 $\um$ filament, suggesting that \nht\ traces the densest gas component in the core. 

Figure 5b, 5c, and 5d show the integrated line emissions of the \nht\ (1,1)-{\bf mg}, -{\bf isg} and (2,2)-{\bf mg}. The \nht\ $(1,1)$-{\bf mg} and -{\bf isg} show similar morphologies, except that the {\bf mg} emission is somewhat saturated due to its higher optical depth. We also used Hyper to extract the compact sources in the \nht\ emission region (Figure 5b). As a result, four gas condensations were found in FIR-3, which are labeled as Cd-1, 2, 3, 4. Another two condensations in FIR-2 are labeled as Cd-n1 and Cd-n2. The physical parameters of the condensations are shown in Table 5. Wherein the \nht\ column densities are estimated using Equation (A4).

Figure 5d shows that the (2,2) emission peak does not coincide with the (1,1) peak but has a noticeable offset of $\simeq3''$ to the northwest. Since $(J,K)=(2,2)$ level has a much higher excitation energy ($E_{\rm u,(2,2)}$=81 K) than (1,1) ($E_{\rm u,(1,1)}$=24 K), the $(2,2)$ emission peak may indicate a higher temperature at its emission peak.

\subsection{The Temperature Distribution from \nht}  
To investigate the temperature distribution from the \nht\ lines, we fitted the line profile using the fiducial radiative transfer model \citep[e.g.][]{friesen09}, which is specified in Appendix A.3. The best-fit line profiles for the spectra at the core center are shown in Figure 4. For the \nht\ $(2,2)$ line, we report the detection of the satellite $F_1=3-2$ line ($f_0=23.721336$ GHz, as labelled on the figure), which has a brightness temperature of $T_{\rm b}=4\pm1$ K. The $(1,1)$ and $(2,2)$ lines are found to have similar excitation temperatures ($T_{\rm ex}=23$ K) and line widths ($\simeq1.0$ \kms), suggesting them tracing the same gas component. 


The \nht\ rotational temperature can then be estimated from the $(1,1)$ and $(2,2)$ emissions using the method in \citet{li13}. From the observed spectra we calculated the intensity ratio between {\bf isg} and {\bf mg}, $R_{\rm sm}=[\int T_{\rm b,isg}(1,1)d\nu]/[\int T_{\rm b,mg}(1,1)d\nu]$, and the intensity ratio between the two transitions, $R_{12}=[\int T_{\rm b}(1,1)d\nu]/[\int T_{\rm b}(2,2)d\nu]$. The rotational temperature is calculated from the two ratios as
\begin{equation}  
T_{\rm rot}=41.5~{\rm K}/\ln[1.06\times C(1,1)\times R_{12}],
\end{equation}
wherein $C(1,1)=0.003+2.26R_{\rm sm}+0.00032\exp(5.38 R_{\rm sm})$ is a correction factor for the (1,1) optical depth. The uncertainty of the $T_{\rm rot}$ is estimated to be $\delta T_{\rm rot}\simeq1.5$ K based on the noise level of the \nht\ lines and the method of \citet[][Equation A.7 therein]{busquet09}. We estimated the relation between $T_{\rm rot}$ and the gas kinetic temperature $T_{\rm k}$ based on the physical conditions in FIR-3 and using the RADEX program \citep{vdtak07}. The calculation is specified in Appendix A.3. It resulted in a nearly thermalized population for $(1,1)$ and $(2,2)$ levels as mainly due to the high density in FIR-3. As a result $T_{\rm rot}$ is close to $T_{\rm k}$. 

As shown in Figure 5e, the $T_{\rm k}$ map has a noticeable peak with $T_{\rm k}\simeq25$ K with the position coincident with the (2,2) emission peak. Away from the peak, $T_{\rm k}$ varies between 15 and 19 K which are significantly lower than the peak value, but similar with the dust temperature in FIR-3. The temperature peak suggests an embedded YSO in condensation Cd-2. Since FIR-3 has no detectable IR point source, it is possible that the YSO is younger than Class-0 and the heating is mainly due to the accretion or core collapse. In this case the $T_{\rm k,peak}$ value may provide some constraint on accretion rate and core evolutionary state. 

We assumed that the central region originally had a similar temperature with its surroundings ($T_{\rm k}=T_{\rm k,avg}=18$ K) in the starless stage, and was then heated to the current value ($T_{\rm k}=T_{\rm k,peak}=25$ K) after the central object started to accrete mass. The luminosity increase at the second stage is then
\begin{equation} 
\Delta L = (4\pi/g) M_{\rm peak}\int k_{\nu}[B_{\nu}(T_{\rm k,peak})-B_{\nu}(T_{\rm
	k,avg})]{\rm d}\nu,
\end{equation}
wherein $M_{\rm peak}$ is the hot-gas mass associated with $T_{\rm k}$ peak. In equation (5) we also assumed that $T_{\rm k}$ can approximate the dust temperature. The $T_{\rm k}$ peak is not resolved thus would represent an average value within one beam, we thus consider the gas mass also within one-beam area, that is $M_{\rm peak}=\pi R_{\rm beam}^2 N({\rm H_2}) \mu m_{\rm H}=0.3~M_\odot$. As the result the luminosity increase is estimated to be $\Delta L \simeq 50~L_\odot$.  

On the other hand, the theoretical accretion luminosity is \citep{stahler80}
\begin{equation} 
L_{\rm acc}=\frac{G M_\ast \dot{M}_{\rm acc}}{R_\ast},
\end{equation}
where $M_\ast$, $R_\ast$, and $\dot{M}_{\rm acc}$ are the mass, radius, and accretion rate for the central object, respectively. The stellar mass can be estimated from its natal core mass assuming a star-forming efficiency of $\epsilon=0.3$ \citep{alves07}. Adopting Cd-2 as the natal gas condensation, we can obtain $M_\ast=\epsilon M_{\rm Cd-2}\simeq 0.3 M_\odot$. We note that the spatial extent of the $T_{\rm k}$ peak may only reflect the spatial range being heated, but is not to define a gas condensation, thus was not used to estimate $M_\ast$. To estimate the accretion rate, we also assume that the accretion energy is eventually released mainly through the dust continuum emission so that $L_{\rm acc}\simeq \Delta L$.


In theory \citep[e.g.][]{young05,tomida10}, the YSO is expected to have $R_\ast=10^2-10^3~R_\odot$ at the FHSC stage and then collapse into $R_\odot$-scale during the protostar formation (second collapse). We derived $\dot{M}_{\rm acc}$ for these two cases. First, if the central object is a first core with $R_\ast=10^2~R_\odot$, the expected accretion rate would be $\dot{M}_{\rm acc} \gtrsim 6\times 10^{-4}~M_\odot$ year$^{-1}$. Alternatively, if the YSO has already collapsed into $R_\odot$-scale, it only needs to have $\dot{M}_{\rm acc}\simeq 6\times10^{-6}~M_\odot$ in order to maintain the observed $\Delta L$. The second $\dot{M}_{\rm acc}$ value is comparable to the average $\dot{M}_{\rm acc}$ for the low-mass protostars \citep{young05} while the first one is two orders of magnitude higher. As an indication, the central YSO would either have a small radius ($\sim 1.0~R_\odot$) or a high accretion rate ($\sim 10^{-4}~M_\odot$ year$^{-1}$) in order to generate the observed $T_{\rm k,peak}$ feature. Correspondingly, the YSO should be either in a rapidly collapsing phase or have lately collapsed into an $R_\odot$-scale protostar. 

\subsection{The Gravitational Instability}  
The possibility for the core collapse can also be evaluated through the Jeans mass, which is
\begin{equation} 
\begin{aligned}
M_{\rm J} & =\frac{\pi}{6}\frac{c_s^3}{G^{3/2}\rho^{1/2}} \\
& = 2 M_\odot \left(\frac{\sigma}{0.2 {\rm km~s^{-1}}}\right)^3\left(\frac{n}{\rm 10^3~cm^-3}\right)^{-\frac{1}{2}}.
\end{aligned}
\end{equation}
$M_{\rm J}$ represents the highest mass that can be sustained by the internal pressure related with the velocity dispersion $\sigma$. FIR-3 has $n\geq10^6$ \cc, and the velocity dispersion $\sigma_{\rm obs}=\Delta V/\sqrt{8\ln(2)}\simeq 0.42$ \kms. These parameters leads to $M_{\rm J}\leq 0.5~M_\odot$ which is smaller than the Cd-2 mass, suggesting that the collapse of Cd-2 is possible unless there is additional support such as magnetic field. Besides FIR-3, FIR-4, 5, and 6 were also suggested to be unstable based on the observed H$_2$CO line widths \citep{watanabe08}. 

Using the \nht\ kinetic temperature, the velocity dispersion due to the thermal motion is estimated to be $\sigma_{\rm th}=k_B T_k/\mu m_{\rm H}=0.25-0.33$ \kms\ as $T_k$ varies between 19 and 27 K. The non-thermal contribution is $\sigma_{\rm nt}=\sqrt{\sigma_{\rm obs}^2-\sigma_{\rm th}^2}\simeq 0.2-0.3$ \kms. The $\sigma_{\rm nt}$ value is in the subsonic rage but tend to be higher than the values in the cold dense gas \citep[e.g. $\sigma=0.1-0.2$ \kms,][]{pineda10}. Based on the current data, it is uncertain whether the FIR cores are experiencing a turbulence decay or oppositely, becoming more turbulent due to the ongoing star-forming activities \citep[e.g.][]{fontani12}. We examined the \nht\ line width over emission region, and found no significant variation. A velocity field over a larger area mapped by single dish may help better reveal the origin of the turbulence.

\section{The Evolutionary State of the FIR Cores}  
\subsection{Mass-and-Luminosity Relation}  
The relation between the luminosity and mass can also provide an estimate for the core evolutionary state \citep{molinari08,elia13}. The Mass-Luminosity diagram is shown in Figure 6. The cores except FIR-4 are quite closely aligned in a power law of $L\propto M^p$ with the index $p=1.0$. The deviation of FIR-4 is expected due to its higher dust temperature. Besides the IRAC source, active star formation in FIR-4 is also indicated by the radio continuum and maser emissions therein \citep{choi15}. FIR-1, 2, and 6 may also have protostars as due to the presence of IRAC sources or outflows \citep{chandler96}. But for these three cores, the stellar emissions should still be too weak to significantly increase the total luminosities. It is also possible that the IR sources at FIR-1 and 2 are just foreground stars. 

Figure 6 also shows the $M-L$ distribution for the other four samples with specified evolutionary stages, including Class-0 protostars \citep{andre00}, prestellar cores \citep{elia13}, starless cores \citep{strafella15}, and recently identified FHSC candidates (with respective references shown in the figure caption). As the figure shows, the more evolved samples do not necessarily have higher absolute luminosities, but instead show steeper slopes for their $M-L$ relations, namely higher $p$ values. The NGC 2024 cores have a similar slope with the prestellar sample ($p=1.0$).

From integration of Equation (1) we can have $L_{\rm core}\propto M_{\rm core} T_{\rm d}^{4+\beta}$. It indicates that luminosity $L_{\rm core}$ can be sensitively affected by $T_{\rm d}$. A sample would have $L\propto M^{1.0}$ only if the cores have quite similar temperatures. If $T_{\rm d}$ varies with the core mass, the $L\propto M$ relation would be different. Assuming a power law of $T_{\rm d}\propto M^q$, we can have $L_{\rm core}\propto M_{\rm core}^{1+q(4+\beta)}$. The starless and prestellar samples exhibit relatively flat power laws, probably because the more massive cores are more shielded from the external heating, leading to $q \leq 0$. In contrast, the protostellar cores might have higher $T_{\rm d}$ in more massive cores ($q>0$) because the massive ones would preferentially become the birthplace for the high-mass stars and/or multiple system. As a result, the relation would become steeper than $L_{\rm core}\propto M_{\rm core}^{1.0}$. The core sample collected from different regions might also be affected by the variation in $\beta$, which could be more complicated and are not discussed here. 

In fact, NGC 2024 would be distinguished from other samples by the presence of strong external heating which have regulated cores into the similar temperatures, and moreover, increased their absolute luminosities to be even higher than the Class-0 sample. In such case, the absolute luminosities of the individual cores would not effectively trace the evolutionary stages. Similarly, the 70 $\micron$ emission would not indicate embedded protostars as it works in cold isolated regions \citep[e.g.][]{pineda11}. In particular, the hot ridge has $T_{\rm d}\simeq55$ K and strong 70 $\micron$ emission, but should be solely illuminated by the external radiation and have no star formation at all. In comparison, as a confirmed protostellar core, FIR-4 only have $T_{\rm d}=22$ K. It could be heated to much higher temperature if located closer to IRS-2b. In this case, the internal emission can be hardly discerned.


\subsection{The Dependence of Dust Temperature on the Stellar Radiation} 
We made a semi-quantitative estimation for the dependence of $T_{\rm dust}$ on the stellar radiation. If IRS-2b has a spectral type of B0V, it would provides an ultraviolet (UV) radiation with photon flux of $N_{FUV}=10^{48}$ s$^{-1}$ \citep[][Chapter 15 therein]{stahler05}. If there is no strong absorbers around the star, the UV radiation at a distance $r$ can be approximated as
\begin{equation}  
\chi(r)=\frac{\overline{E} N_{FUV}}{4\pi r^2}.
\end{equation} 
We adopted $\overline{E}=9$ eV as the average photon energy for the UV radiation field. For a dust core with high opacity in optical ($A_V\gg 1$) in thermal equilibrium, the dust temperature would be \citep{li03}
\begin{equation}  
T_{\rm d}=12.0\times \left[\frac{\chi(r)}{\chi_0}\right]^{1/6}~{\rm K},
\end{equation}
where $\chi_0=2.0\times10^{-4}$ ergs s$^{-1}$ cm$^{-2}$ sr$^{-1}$ is the unit 
intensity for the interstellar radiation field \citep{draine78}. With the two equations joined together, $T_{\rm d}$ would be related with $N_{FUV}$ and $r$ in the form 
\begin{equation}  
\begin{aligned}
T_{\rm d}(r) & =12.0 \left(\frac{\overline{E} N_{FUV}}{4\pi \chi_0}\right)^{1/6} r^{-1/3} \\           
& =110~{\rm K} \times \left(\frac{\overline{E}}{9~{\rm eV}}\right)^\frac{1}{6} \times \left(\frac{N_{FUV}}{10^{48}~ {\rm s^{-1}}}\right)^\frac{1}{6} \times \left(\frac{r}{\rm 0.1~pc}\right)^{-\frac{1}{3}}.
\end{aligned}
\end{equation}
The modelled $T_{\rm d}(r)$ function is shown in Figure 7. The $T_{\rm d}(r)$ curves for B1 and O9 stars are also presented for comparison, with the range between them highlighted in yellow area. The data for the FIR cores and the five positions on the ridge are presented on the diagram. Their $r$ values represent the projected distances from IRS-2b. 

At the observed distances, the model predicts a range of $T_{\rm d}\simeq55-70$ K, which reasonably agrees with the temperature on the hot ridge, but is completely deviated from the core temperatures (17-22 K). The low temperatures in the FIR cores may indicate the actual distances from IRS-2b being much larger than the projected values as seen on the image. In comparison, the hot ridge should have a small uncertainty in its distance because it mainly traces the PDR near IRS-2b. The data points for the ridge suggests that the modeled $T_{\rm d}(r)$ function for B0 star can reasonably describe the dust heating in NGC 2024. Based on the model, the temperatures of FIR cores would indicate a distance of $r\simeq5$ pc. As a result, the filament should be well separated from the hot region around IRS-2b, especially considering the fact that the hot gas nebula only has a spatial extent of 0.2-0.4 pc as shown in the MIPS 24 $\um$ emission. 

In the case that the intervening absorption between IRS-2b and the filament is not negligible, the U-V radiation would be transferred to longer wavelengths. In this case the $T_{\rm d}(r)$ profile can be evaluated using another dust heating model \citep{scoville76,garay99,wang11}:
\begin{equation} 
T_{\rm d}(r)=71\left(\frac{\rm 0.1~pc}{r}\right)^{2/(4+\beta)}\left(\frac{L_{\rm star}}{10^5~L_\odot}\right)^{1/(4+\beta)}\left(\frac{0.1}{f} \right)^{1/(4+\beta)},
\end{equation}
where the dust opacity index is adopted as the value for the hot component ($\beta=1.6$). The luminosity is adopted as the total luminosity of the main-sequence B0 star that is $L=5\times10^5~L_\odot$ \citep{stahler05}. $f=0.08$ cm g$^{-1}$ is the dust emissivity at $\lambda=50~\micron$ for a condition of $\tau_{100\um}=0.1$ and $n\sim 10^4$ \cc\ \citep{scoville76}. The derived $T_{\rm d}(r)$ relation is plotted with thick gray line in Figure 7. This $T_{\rm d}(r)$ curve is only slightly bellow the UV radiation model for the B0 star, and is also suggestive of a large distance for the FIR cores, i.e. 2 to 3 pc from IRS-2b.    

As another possibility for the low temperature, the filament can have a strong self-shielding due to its higher opacity, thus can also maintain a low temperature even if located close to IRS-2b. However, if the filament is near the ridge, its outer layer would also be heated to a similar temperature with the ridge, and then exhibit a hot component in its SED \textit{in addition} to the cold one. In fact, the FIR cores are much fainter than the ridge at $\lambda \leq 100~\um$, and their SEDs are completely dominated by the cold component. We thus conclude that the filament should indeed have a large distance from IRS-2b. The previous studies showed that the dense molecular gas should be located behind the H{\sc ii} region and PDR \citep{roshi14}. The current data further shows that the hot and cold components are actually spatially separated rather next to each other. The expected spatial layout for the filament and hot gas around IRS-2b are shown in Figure 7b.


\section{Summary}   
We present an observational study with dust continuum and high-resolution \nht\ line emissions to reveal the internal structures and physical properties of the FIR cores in NGC 2024 filament. A particular focus of this paper is the most massive core FIR-3 therein. The main results are:

1. All the FIR cores in the filament are found to have low dust temperatures between 17.5 and 22 K. FIR-3 has a compact morphology, dust temperature of 17.5 K, and no significant IR sources. The physical properties are suggestive of an evolutionary stage younger than Class-0.

2. In FIR-3, the \nht\ line emissions exhibit a compact temperature rise feature with a peak value of $T_{\rm k}=25$ K above the surrounding level of $T_{\rm k}=15-19$ K. From the $T_{\rm k}$ peak, the central YSO is estimated to have an accretion luminosity of $\Delta L \sim50~L_\odot$. The temperature rise without infrared counterpart suggests an ongoing or lately occurred collapse from FHSC to Class-0 protostar. 

3. The FIR cores in NGC 2024 filament have comparable or even higher luminosities than the Class-0 cores in the same mass range, but mostly due to the external heating from IRS-2b. On the mass-luminosity diagram, the FIR cores (except FIR-4) exhibit a correlation with the power law index ($p=1.0$) comparable to the prestellar sample. For each core, the internal stellar heating appears weak and has no significant influence on temperature or luminosity.

4. The FIR cores have much lower dust temperatures than the hot ridge (56 K). The difference in their $T_{\rm d}$ suggests that the filament should be much more distant from the central heating star IRS-2b. Comparison with the dust heating models shows that the filament should be separated from the hot ridge by $D=5$ pc along the line of sight. The filament and ridge also have distinct dust opacity indices of $\beta \simeq 2.7$ and 1.6, respectively.       

The observed physical properties suggest an overall cold dense state in the filament. And FIR-3 could be an intermediate-mass core with an YSO possibly at a transition stage from the FHSC to Class-0. Further high-resolution observations (e.g. with ALMA) are expected to resolve its internal structures and reveal the details of the collapse and the accretion state. 

\section*{Acknowledgment}
We express our gratitude to Dr. A. Traficante for kindly providing a detailed instruction for the usage of Hyper. We thank the anonymous referee for the careful inspection of the manuscript and constructive comments. This work is supported by the China Ministry of Science and Technology under State Key Development Program for Basic Research (973 program) No. 2012CB821802, the National Natural Science Foundation of China No. 11403041, No. 11373038, No. 11373045, Strategic Priority Research Program of the Chinese Academy of Sciences, Grant No. XDB09010302, and the Young Researcher Grant of National Astronomical Observatories, Chinese Academy of Sciences. This research has made use of "Aladin sky atlas" developed at CDS, Strasbourg Observatory, France.
\clearpage

\begin{table*}
{\scriptsize
\centering
\begin{minipage}{140mm}
\caption{The positions and sizes of the of the cores as fitted by Hyper.}
\begin{tabular}{llcccccc}
\hline\hline
Object      &  position at 450 $\um$    &  $70\um$ offset$^a$   &  $160\um$ offset$^a$      &  $r_{\rm maj}^b$  &  $r_{\rm min}^b$    &  pa$^b$     &   $r_{\rm core}^c$    \\
\quad       &  (RA,Dec)                 &  (arcsec)             &  (arcsec)                 &  (arcsec)         &  (arcsec)           &  (degree)   &   (arcsec)           \\
\hline 
FIR-1       &  05:41:41.40 -01:53:47.0  &  unidentified         &  (1.4,-3.6)               &  12               &  11                 &  126        &   10/8.9  \\
FIR-2       &  05:41:42.50 -01:54:06.0  &  unidentified         &  (0.7,-1.8)               &  16               &  9                  &  157        &   9/7.8  \\
FIR-3       &  05:41:43.00 -01:54:24.0  &  unidentified         &  (-0.4,-2.4)              &  15               &  10                 &  146        &   10/8.9  \\
FIR-4       &  05:41:44.00 -01:54:42.0  &  (0.36,-4.68)         &  (0.4, -4.3)              &  13               &  9                  &  147        &   9/7.8  \\
FIR-5a      &  05:41:44.10 -01:55:40.0  &  (1.8,8.64)           &  (2.9,-2.5)               &  14               &  10                 &  269        &   11/10.0  \\
FIR-5b      &  05:41:44.80 -01:55:32.0  &  (0.36,-2.16)         &  (0.4,-4.7)               &  13               &  10                 &  107        &   12/11.1  \\
FIR-6       &  05:41:45.00 -01:56:01.0  &  (1.44,-2.16)         &  (1.8,-2.4)               &  16               &  11                 &  157        &   12/11.1  \\
FIR-7       &  05:41:44.90 -01:56:15.0  &  unidentified         &  (2.5,6.5)                &  16               &  9                  &  184        &   10/8.9  \\
\hline
\end{tabular} \\
$a.${ The RA and Dec offset at 70 and 160 $\um$ bands relative to the core center positions at 450 $\um$. The cores unidentified by Hyper are not shown. }  \\
$b.${ The major, minor axes, and position angles (pa). The pa values are counter-clockwise to the north. }  \\
$c.${ The best-fit core radius assuming a circular shape for the core area at 450 $\um$. The second value is the radius deconvolved with the beam size, i.e. $r_{\rm core}^2=r_{\rm obs}^2-(\theta_{\rm FWHM}/2)^2$, wherein $\theta_{FWHM}=9''$ is the FWHM beam size. }  \\
\end{minipage}
}
\end{table*}

\begin{table*}
{\scriptsize
\centering
\begin{minipage}{140mm}
\caption{The maginitude of the IR sources associated with the cores in IRAC bands and the flux densities of the cores in MSX bands. }
\begin{tabular}{lccccccc}
\hline\hline
Object     &  $M_{\rm 3.6\um}^a$  &  $M_{\rm 4.5\um}^a$  &  $M_{\rm 8.0\um}^a$   &  $S_{8\um}$   &  $S_{12\um}$   &  $S_{15\um}$   &  $S_{21\um}$  \\
\quad      &  (mag)               &  (mag)               &  (mag)                &  (Jy)         &  (Jy)          &  (Jy)          &  (Jy)         \\
\hline                                                                             
FIR-1      &  9.09                &  8.48                &  N                    &   $<0.2$      &  $<0.5$        &  $<0.6$        &  $<3.0$       \\        
FIR-2      &  9.90                &  9.17                &  N                    &   $<0.2$      &  $<0.5$        &  $<0.6$        &  $<3.0$       \\             
FIR-3      &  N                   &  N                   &  N                    &   $<0.2$      &  $<0.5$        &  $<0.6$        &  $<3.0$       \\            
FIR-4      &  8.69                &  7.09                &  N                    &   $<0.2$      &  $<0.5$        &  $<0.6$        &  $<3.0$       \\            
FIR-5$^b$  &  N                   &  N                   &  N                    &   43          &  50            &  65            &  325          \\             
FIR-6      &  N                   &  N                   &  N                    &   $<0.2$      &  $<0.5$        &  $<0.6$        &  $<3.0$       \\          
FIR-7      &  N                   &  N                   &  N                    &   $<0.2$      &  $<0.5$        &  $<0.6$        &  $<3.0$       \\     
Ridge$^c$  &  N                   &  N                   &  N                    &   37          &  32            &  35            &  307          \\  
\hline                                                                                                      
\end{tabular} \\
$a.${ The IRAC magnitudes in 3.6, 4.5 and 8.0 $\um$ bands for the point source possibly associated with the FIR cores, from Megeath et al. (2012). The label 'N' means non-detection, and would indicate $M_{3.6}<14.5$. } \\
$b.${ FIR-5a and 5b are blended, thus the measured flux density should represent a total value of the two. The emission from the hot ridge is also included. } \\
$c.${ In the IRAC 8 $\um$ and all the MSX bands, the flux density is an averaged value for the four box regions shown in Figure 3b. The four regions are individually examined and found to have small differences in $T_{\rm d}$ ($<\pm2$ K).} \\
\end{minipage}
}
\end{table*}

\begin{table*}
{\scriptsize
\centering
\begin{minipage}{140mm}
\caption{The continuum flux densities (in unit of Jy) of the cores and the ridge in Herschel and SCUBA wavebands.}
\begin{tabular}{lcccccccc}
\hline\hline
Object      &  $S_{70\um}$  &  $S_{100\um}$  &  $S_{160\um}$   &  $S_{250\um}$   &  $S_{350\um}$    &  $S_{500\um}$   &  $S_{450\um}$  & $S_{850\um}$ \\
\hline                                            
FIR-1       &   130         &  285           &  683            &  295           &  121             &  49        &  65       &  3.5     \\        
FIR-2       &   304         &  650           &  873            &  472           &  205             &  58        &  101       &  5.5     \\             
FIR-3       &   330         &  743           &  1282           &  694           &  298             &  80        &  156      &  7.0     \\            
FIR-4       &   1133        &  1620          &  1260           &  577           &  232             &  76        &  107       &  4.5     \\            
FIR-5$^a$   &   3571        &  3655          &  3951           &  1918          &  625             &  125       &  125/97    &  12.5    \\             
FIR-6       &   313         &  660           &  922            &  646           &  268             &  58        &  111      &  6.0     \\          
FIR-7       &   165         &  442           &  804            &  479           &  168             &  49        &  75       &  4.0     \\     
Ridge$^b$   &   2376        &  1182          &  497            &  141           &  37              &  10        &  15       &  1.7     \\  
\hline                                                                                            
\end{tabular} \\
$a.${ FIR-5a and 5b are blended except at 450 $\micron$. The $S_{450\um}$ values for the two cores are separately given, while $S_{\lambda}$ at other bands represent the total value of the two. At $\lambda<100~\um$, the emission from the hot ridge is significant and would also contribute to the total $S_\lambda$.} \\
$b.${ In each band, the flux density represents an average value for the four box regions as shown in Figure 3b. The four regions are individually examined and found to have a small difference of $\Delta T_{\rm d}$ ($<2$ K).} \\
\end{minipage}
}
\end{table*} 

\begin{table*}
{\scriptsize
\centering
\begin{minipage}{140mm}
\caption{The physical parameters of the cores and the ridge.}
\begin{tabular}{lcccccc}
\hline\hline
Object      &  $T_{\rm dust}$  &  $\beta$       &   $N_{\rm tot}$              &   $n^c$                    &  mass              &  $L_{\rm bol}$     \\
\quad       &  (K)             &  \quad         &   (cm$^{-2}$)                &   (cm$^{-3}$)              &  ($M_\odot$)       &  ($L_{\odot}$)  \\  
\hline                                                                                                                   
FIR-1       &  $18.5(1.0)$     &  $2.5(0.3)$    &   $3.5(0.7)\times10^{23}$    &   $3.1(0.9)\times10^6$     &  $4.5(0.9)$        &  80 (40)            \\
FIR-2       &  $18.0(1.0)$     &  $2.7(0.3)$    &   $4.3(0.8)\times10^{23}$    &   $2.4(0.7)\times10^6$     &  $5.2(3.0)$        &  130(70)           \\
FIR-3       &  $17.5(1.0)$     &  $2.6(0.3)$    &   $7.4(0.9)\times10^{23}$    &   $1.2(0.3)\times10^7$     &  $10.0(4.0)$       &  220(80)           \\
FIR-4       &  $22.0(1.0)$     &  $2.9(0.3)$    &   $3.6(0.5)\times10^{23}$    &   $6.0(1.8)\times10^6$     &  $5.0(2.0)$        &  570(100)          \\
FIR-5a$^a$  &  $17.5(1.0)$     &  $2.7(0.3)$    &   $4.0(0.6)\times10^{23}$    &   $3.2(0.9)\times10^6$     &  $5.9(3.0)$        &  120(50)           \\
FIR-5b      &  $17.5(1.0)$     &  $2.7(0.3)$    &   $3.7(0.5)\times10^{23}$    &   $2.6(0.7)\times10^6$     &  $4.5(2.0)$        &  90 (40)            \\
FIR-6       &  $18.5(1.0)$     &  $2.7(0.3)$    &   $3.8(0.5)\times10^{23}$    &   $1.8(0.4)\times10^6$     &  $5.8(2.0)$        &  160(40)           \\
FIR-7       &  $18.0(1.0)$     &  $2.6(0.3)$    &   $3.6(0.5)\times10^{23}$    &   $1.8(0.4)\times10^6$     &  $4.9(1.5)$        &  110(30)           \\
ridge$^b$   &  $56(1.0)$       &  $1.6(0.2)$    &   $3.1(0.6)\times10^{22}$    &   $2.5(0.8)\times10^5$     &  $0.5(0.2)$        &  600(300)          \\
\hline
\end{tabular} \\
$a.${ The dust temperature is fitted from the total flux densities for FIR-5a and 5b.} \\
$b.${ Each physical parameter represents an average for the four square regions as shown in Figure 3b. } \\
$c.${ For the dense cores, the density is derived from the peak total column density and the mean radius. For the ridge, the density is derived from the column density assuming the thickness is equal to the average width.} \\
\end{minipage}
}
\end{table*}

\begin{table*}
{\scriptsize
\centering
\begin{minipage}{140mm}
\caption{The physical parameters of the \nht\ gas condensations.}
\begin{tabular}{lcccccc}
\hline\hline
Object      &  offset$^a$          &  $T_{\rm k}$     &  radius      &   mass$^b$     &  $N$(\nht)  \\
\quad       &  (arcsec)            &  \quad             &  (arcsec)    &  ($M_\odot$)   &  ($10^{15}$ cm$^{-2}$) \\
\hline                                                                                         
Cd-1        &  (-3.0,7.0)          &  18(2)        &  3.5(0.5)       &  1.2(0.8)      &  3.0  \\    
Cd-2        &  (-1.5,4.0)          &  25(2)        &  3.5(0.5)       &  1.0(0.7)      &  2.9  \\
Cd-3        &  (0,0)               &  19(2)        &  4.0(0.5)       &  2.0(1.5)      &  3.8  \\
Cd-4        &  (2.5,-3.5)          &  17(2)        &  4.0(0.5)       &  0.8(0.5)      &  2.7  \\
Cd-n1       &  (-7.5,20.5)         &  16(2)        &  3.5(0.5)       &  0.6(0.4)      &  2.3  \\
Cd-n2       &  (-6.5,14.5)         &  14(2)        &  3.5(0.5)       &  0.3(0.2)      &  1.7  \\
\hline
\end{tabular} \\
$a.${ The RA and DEC offset from the Cd-3. Cd-3 is centered at RA=05:35:26.53, Dec=-05:04:00.6. The condensations are shown in Figure 5b.} \\
$b.${ The Cd-3 mass is calculated assuming its average column density is equal to the peak total column density in FIR-3 (Table 4), i.e. $M=\pi R^2 \mu m_0 N_{\rm tot,FIR3}$. The masses of the other condensations are calculated by comparing with Cd-3 and assuming that mass is proportional to \nht\ (1,1)-{\bf isg} integrated intensity, i.e. $M_{\rm Cd\textendash i}=M_{\rm Cd\textendash 3}(S_{\rm Cd\textendash i}/S_{\rm Cd\textendash 3})$. } \\
\end{minipage}
}
\end{table*}         

\begin{figure*}
\centering
\includegraphics[angle=0,width=1.0\textwidth]{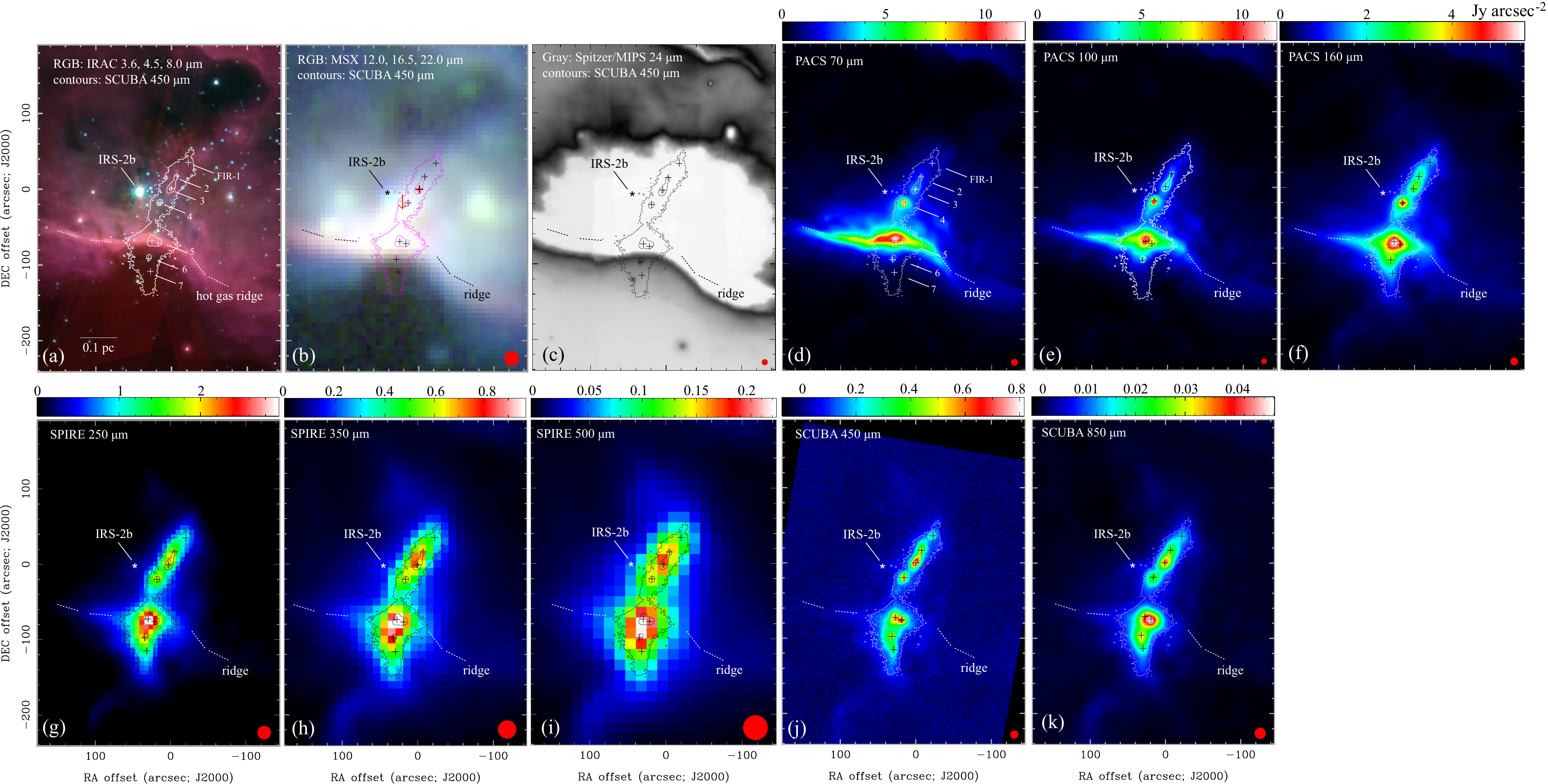} \\
\caption{\small The continuum emissions in the NGC 2024 filament region. {\bf (a)} The RGB image of the IRAC 3.6, 4.5, and 8.0 $\micron$ bands. {\bf (b)} The RGB image of the MSX 12, 16.5, and 21 $\micron$ bands. {\bf (c)} The MIPS 24 $\micron$ image, wherein the blank white region is saturated. {\bf (d)-(k)} the images in Herschel PACS 70, 100, and 160 $\um$ bands, SPIRE 250, 350, and 500 $\um$ bands, JCMT/SCUBA 450 and 850 $\micron$ bands. For the Herschel and SCUBA images, the intensity units are scaled to Jy arcsec$^{-2}$ without changing the image resolution or pixel size. In each panel, the SCUBA 450 $\um$ emission is plotted in contours. The contour levels are 10, 50, 90\% of the peak value (0.81 Jy arcsec$^{-2}$). And the locations of IRS-2b and hot gas ridge are labeled. \label{fig1}}
\end{figure*}

\begin{figure*}
\centering
\includegraphics[angle=0,width=0.5\textwidth]{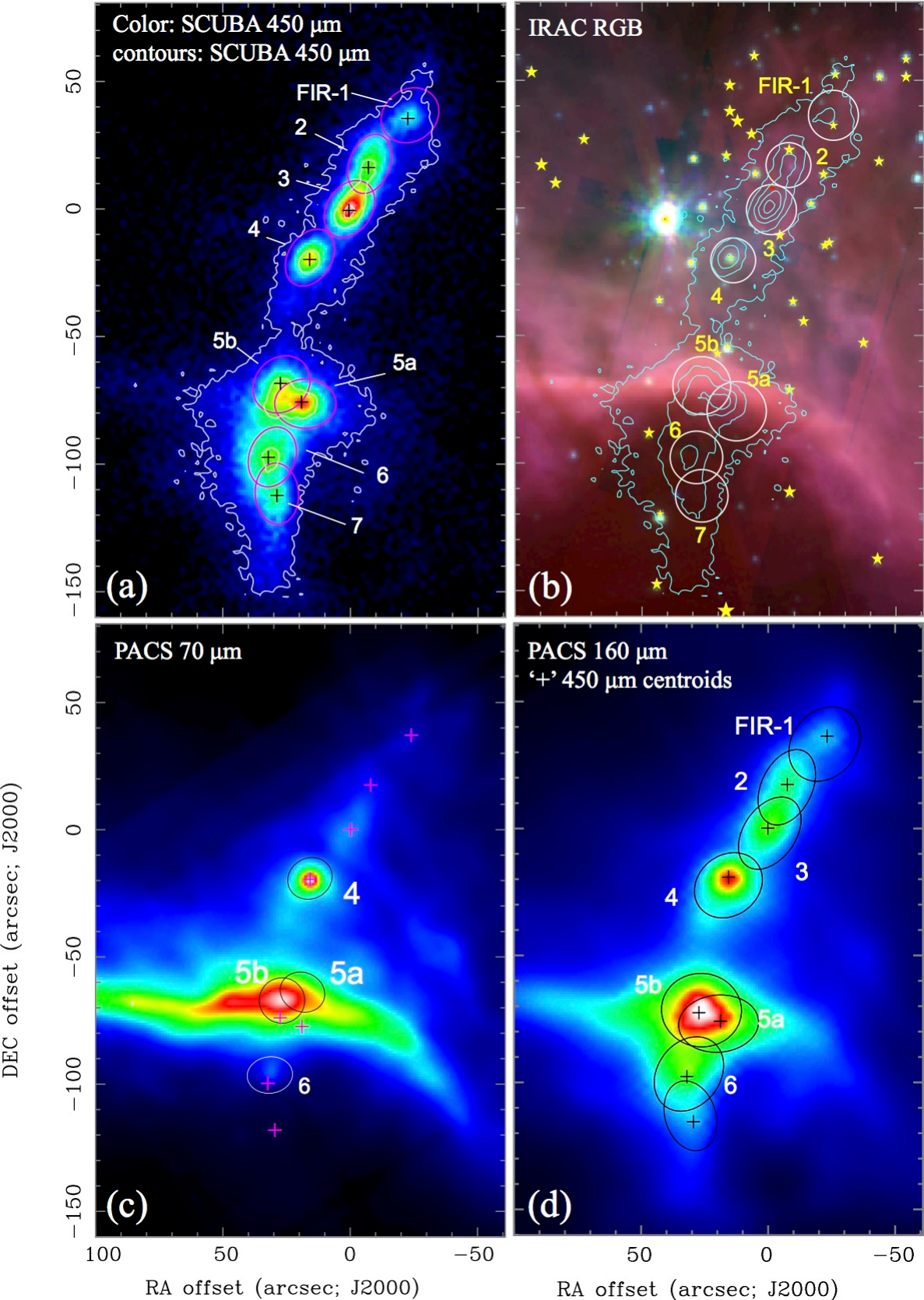} \\
\caption{\small The dense core extraction on the {\bf (a)-(b)} 450, {\bf (c)} 70, and {\bf (d)} 160 $\um$ images using the Hyper routine. The 450 $\micron$ contours are 10, 30, 50, 70, 90\% of the peak value. In panel {\bf (a)}, each ellipse indicates the core area fitted by Hyper, with the semi-axes equal to the FWHMs in 2D Gaussian fit. In panel {\bf (b)}, the fitting is performed assuming each core to have a circular shape. The 450 emission contours and the cores overlaid on the IRAC RGB image. The yellow stars indicate the protostars identified in the IRAC bands \citep{megeath12}. We note that at 70 $\um$, the Core FIR-5 is blended with the hot dust emission, thus the two sources should represent the emission peaks rather than dense cores. \label{fig2}}
\end{figure*}

\begin{figure*}
\centering
\includegraphics[angle=0,width=1.0\textwidth]{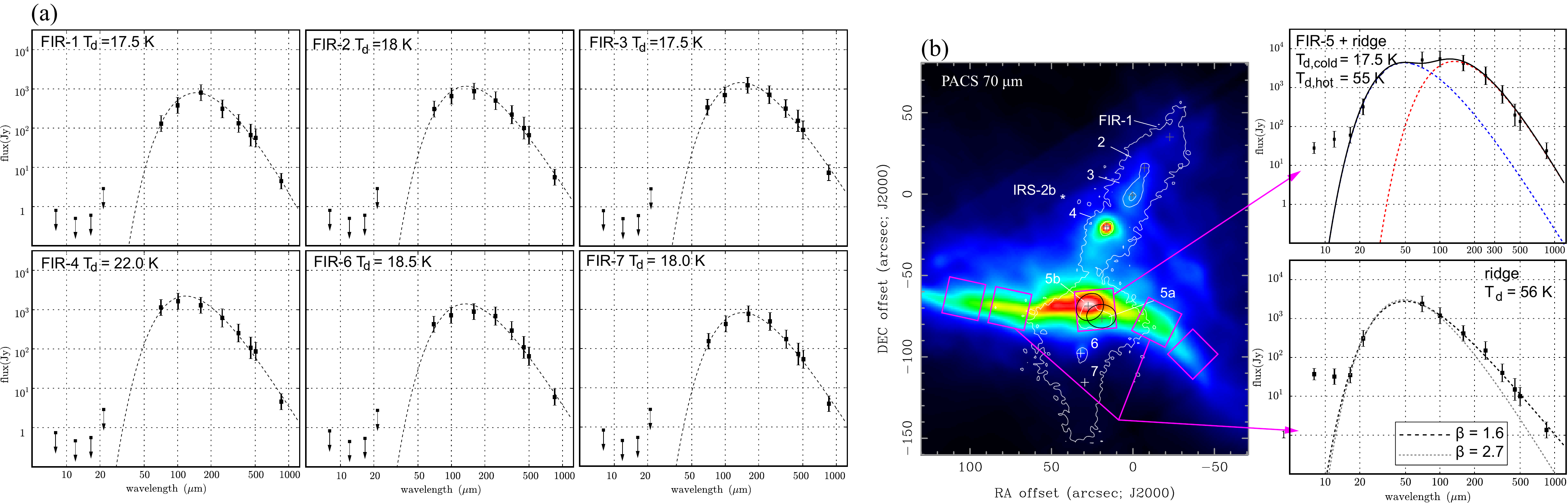} \\
\caption{\small {\bf (a)} The spectral energy distributions (SEDs) for the dense core FIR-1, 2, 3, 4, 6, 7. In each panel, the black square with error bars are the observed values, and the dashed line is the best-fit SED profile. The cores are not clearly detected above the surrounding extended emissions in the IRAC and MSX bands, the data points are thus shown with downward arrows to indicate upper limits. {\bf (b)} The SEDs for the five areas on the ridge. At each position, the flux densities are measured within a $25''$-wide square. The upper-right panel shows the SED at the central position that contains the blended emission from FIR-5. And an SED with two components are used to fit the data. The lower panel shows the averaged flux densities of the other four positions. The dotted lines show the best-fit SED for 20 to 100 $\micron$ data using a fixed value of $\beta=2.7$, in order to exhibit the deviation from the best-fit SED at longer wavelengths. The MSX 16.5 and 21 $\micron$ data are also adopted in fitting the hot dust component. The MSX 12 $\micron$ intensity is above the SED curve and may indicate component with even higher temperature, while the IRAC 8 $\micron$ emission may largely represent the PAH component. \label{fig3} }
\end{figure*}

\begin{figure*}
\centering
\includegraphics[angle=0,width=0.5\textwidth]{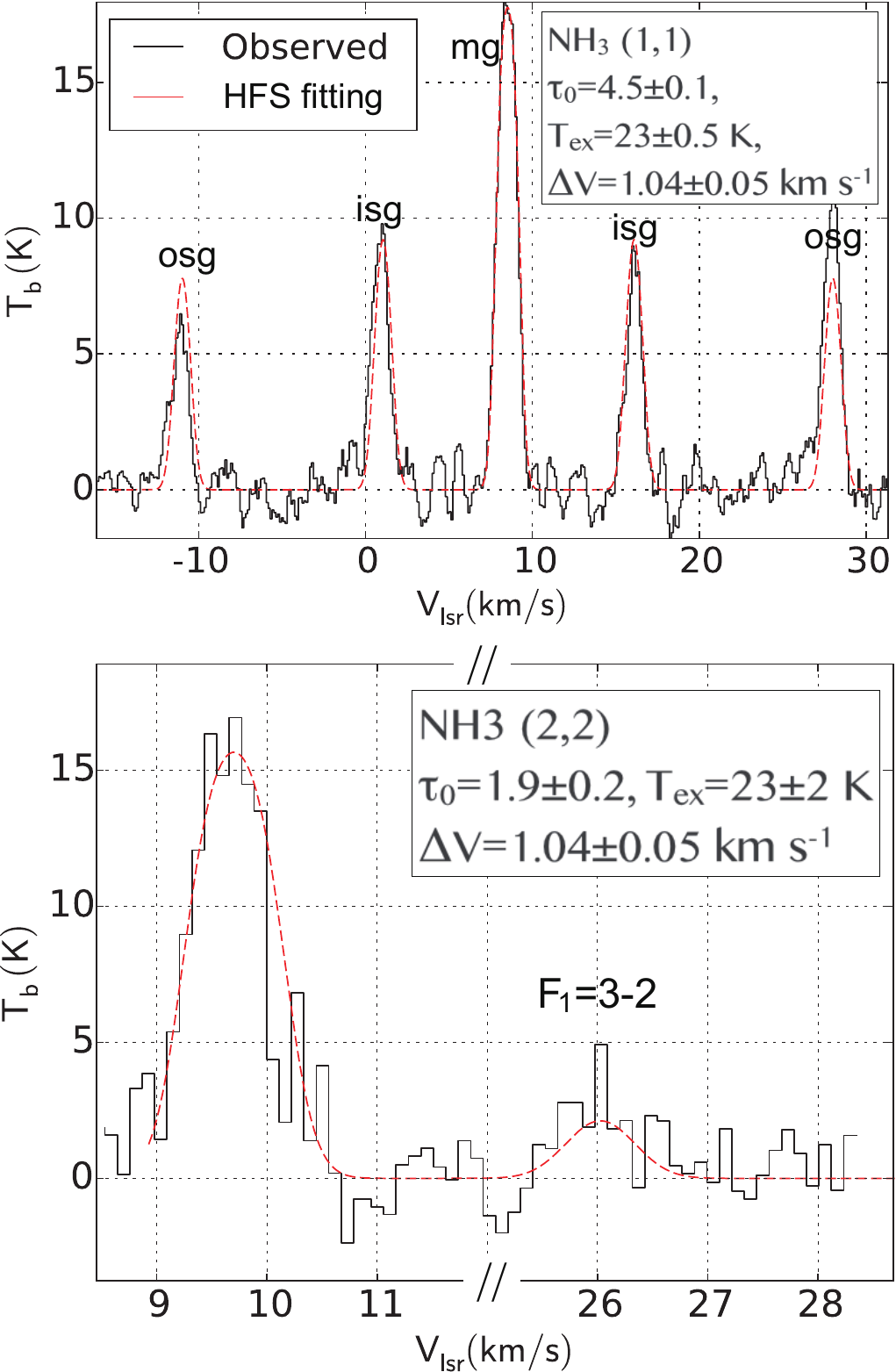} \\
\caption{\small The VLA \nht\ (1,1) and (2,2) spectra at the FIR-3 center. In each panel, the black solid line and the red dashed line represent the observed spectrum and the HFS fitting, respectively. The HFS is denoted on each emission peak, wherein "mg", "isg", and "osg" represent the main, inner-satellite, and outer-satellite groups, respectively. \label{fig5}}
\end{figure*}

\begin{figure*}
\centering
\includegraphics[angle=0,width=1.0\textwidth]{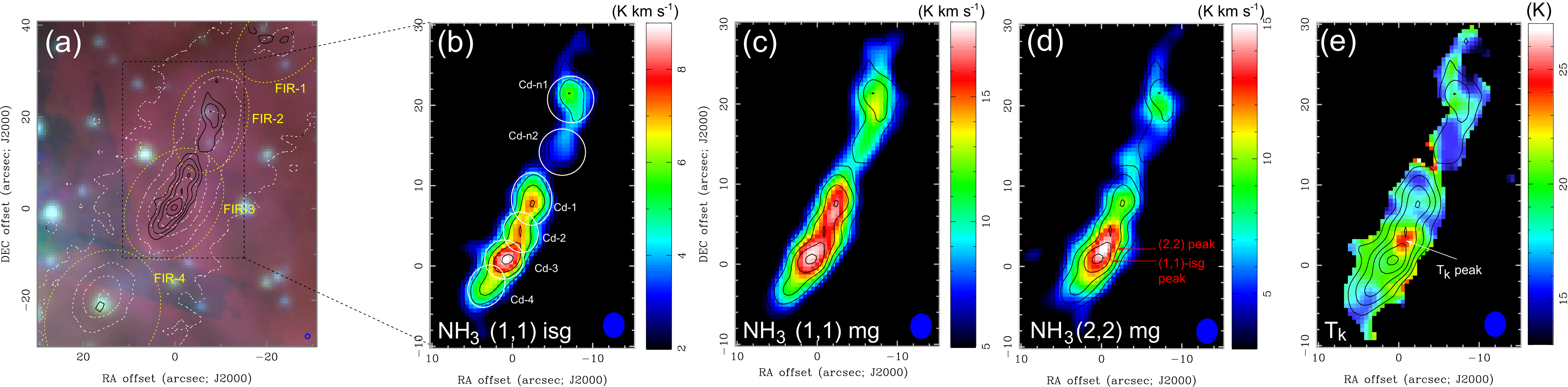} \\
\caption{\small {\bf (a)} The integrated \nht\ (1,1) isg emission (black contours) overlaid on the IRAC RGB image; the SCUBA 450 $\um$ emission is also presented in white contours. The levels are 10, 30, 50, 70, 90\% of the maximum. The yellow dashed ellipses denote the FIR core areas fitted by Hyper. {\bf (b)} The integrated \nht\ (1,1)-isg emission (false color image and contours). The contour levels are 4, 8, 12, 16, 20 times the rms level (0.6 K \kms). The white circles indicate the condensation areas fitted by Hyper. {\bf (c)} The (1,1)-isg emission (contours) overlaid on the (1,1)-mg emission (false-color image). {\bf (d)} The (1,1)-isg emission (contours) overlaid on the integrated (2,2) emission (false-color image). {\bf (e)} The $T_{\rm k}$ map estimated from the (1,1) and (2,2) emissions. The contours are also (1,1)-isg emission. \label{fig6}}
\end{figure*}

\begin{figure*}
\centering
\includegraphics[angle=0,width=0.7\textwidth]{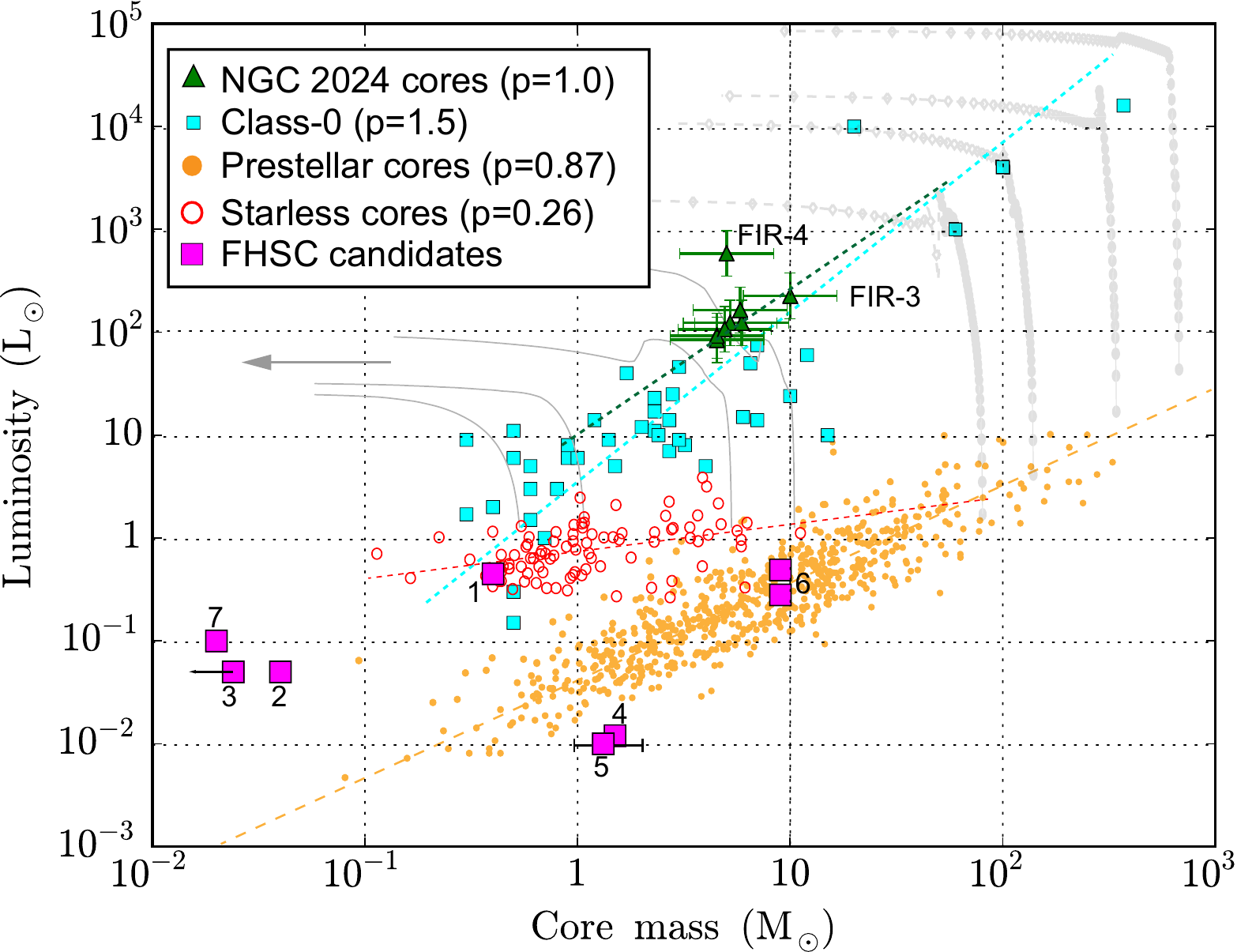} \\
\caption{\small The bolometric luminosity as a function of core mass for the NGC 2024 cores (green triangles), Class-0 protostars (cyan squares, Andr\'{e} 2000), FHSC candidates (purple squares), prestellar cores (orange dots, Elia et al. 2013), and star less cores (red circles, Strafella et al. 2015). The dashed lines represent the best-fit power-law relation for each sample. For NGC 2024, FIR-4 is excluded from the relation of other cores. The solid and solid-dotted grey lines represent the evolutionary tracks for the low- and high-mass YSOs, with the arrow indicating the evolution direction. The FHSC candidates are from: $^{1}$\citet{belloche06}, $^{2}$\citet{chen10}, $^{3}$\citet{pineda11}, $^{4}$\citet{enoch10}, $^{5}$\citet{dunham11}, $^{6}$\citet[][two cores]{pezzuto12}, $^{7}$\citet{friesen14}. The numbers for the literatures are labelled with the data points.}
\end{figure*}

\begin{figure*}
\centering
\includegraphics[angle=0,width=0.9\textwidth]{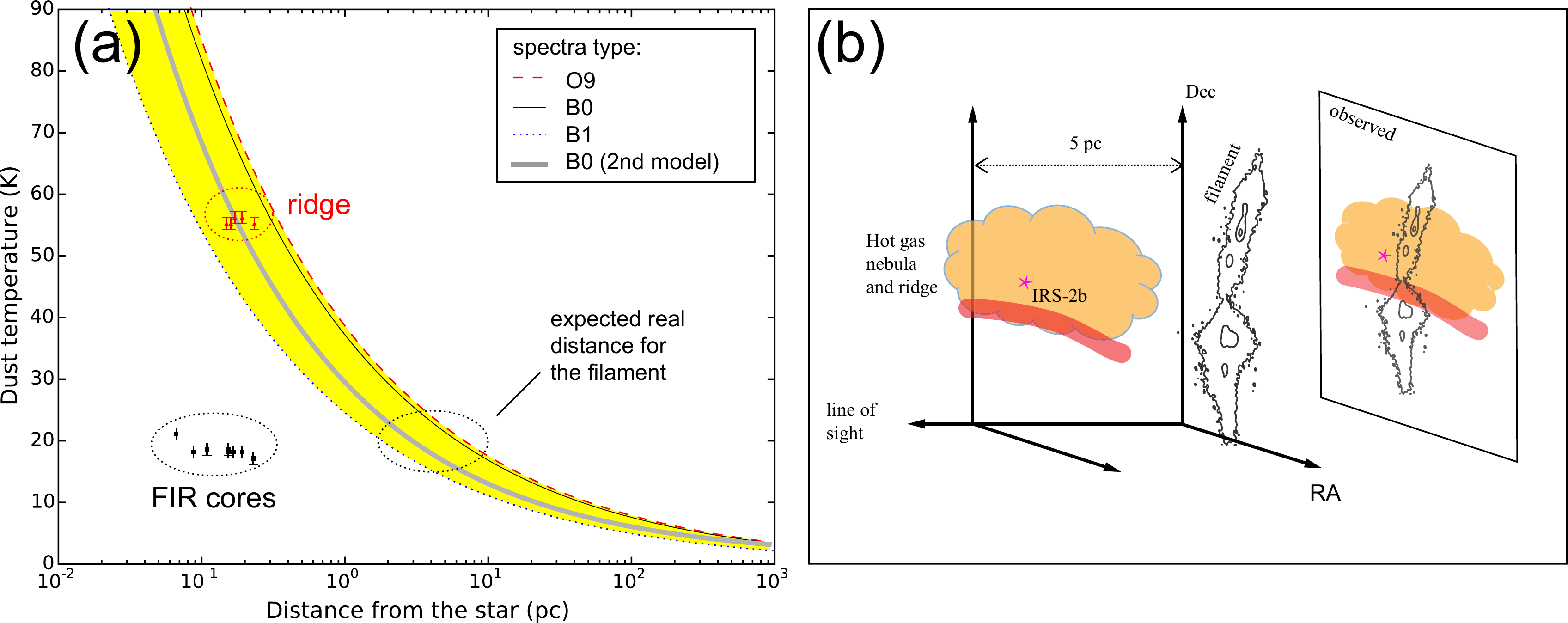} \\
\caption{\small {\bf (a)} the dust temperature $T_{\rm d}$ under external radiation field of a central heating star as a function of the stellar distance $r$ \citep{li03}. The $T_{\rm d}(r)$ curves corresponding to the three different stellar types are plotted with different lines for comparison. The "second model" refers to that of Scoville \& Kwan (1976), with the relation expressed as Equation (11). The $(T_{\rm d},r)$ data measured for the hot ridge and the FIR cores in the filament are shown. For the data points, $r$ represents the projected distance from IRS-2b. {\bf (b)} A schematic view showing the spatial layout of the hot-gas/ridge and the filament. }    
\end{figure*}

\clearpage


\appendix

\section{Appendix material}
\renewcommand\thetable{A\arabic{table}}
\setcounter{table}{0}
\renewcommand\thefigure{A\arabic{figure}}
\setcounter{figure}{0}

\subsection{Using Hyper to measure the flux densities}
The flux densities of the FIR cores are measured using Hyper \citep{traficante15}. In each band, the core radius is convolved with the beam size, namely $r_{\rm conv}=\sqrt{r_{\rm core}^2+r_{\rm beam}^2}$, in order to rightly cover the observed core areas. These parameters are adopted from the 450-$\um$ fitting results because the 450-$\um$ band ideally trace the cold dense cores. Hyper can subtract the background emission and separate the overlapped cores using multi-gaussian fitting. 

The flux errors are mainly determined by the three factors: (1) flux calibration, (2) background emission subtraction, and (3) core area, i.e. aperture size. The flux calibration has an uncertainty of 10\% for PACS\footnote{\scriptsize see PACS observer's manual: http://herschel.esac.esa.int\\/Docs/PACS/html/pacs\_om.html} and 5\% for SPIRE\footnote{\scriptsize see SPIRE observer's manual: \\ http://herschel.esac.esa.int/twiki/\\bin/view/Public/SpireCalibrationWeb}, 30\% for SCUBA 450 $\um$ and 10\% for SCUBA 850 $\um$ \citep{johnstone99}. The background fitting depends on its polynomial order. Varying the order from 0 to 3 would cause an uncertainty of $\sim10\%$ for the flux density. The core area depends the sigma threshold adopted in the fitting. As it varies from 4 to 6 $\sigma$, the core size would vary by $\sim30\%$, causing the flux density to vary by $\sim40\%$. Considering these factors, the flux error would be $\sim70\%$ for PACS, $\sim60\%$ for SPIRE, $\sim100\%$ for SCUBA 450 $\um$ and $\sim70\%$ for SCUBA 850 $\um$. The flux densities in the Herschel bands are also corrected for the photometric colour-correction factors (CC). 

The CC is a function of the dust temperature \footnote{For PACS, see \scriptsize M\"{u}ller, Okumura \& Klaas, \textit{PACS Photometer - Colour Corrections}, http://herschel.esac.esa.int/twiki/pub/Public/PacsCalibrationWeb\\/cc\_report\_v1.pdf. For SPIRE, see \textit{SPIRE hand book}, http://herschel.esac.esa.int/Docs/SPIRE/html/spire\_om.html. }. The correction includes two steps, which are performed iteratively: 1) doing the SED fitting to determine the dust temperature for each core (see Section 4) and the related CC; 2) using CC to correct the flux density of each core. The temperature variation due to CC is small. The $T_{\rm d}$ values become converged within two iterations. As shown in Section 4.1, the SED fitting reveals two components with $T_{\rm d}\simeq55$ K and $T_{\rm d}\simeq17.5$ K. The cold component (FIR cores except FIR-5) has CC$\_70=1.269$, CC$\_100=1.051$ and CC$\_160=0.964$, and the hot component (ridge) has CC$\_70=0.982$, CC$\_100=0.982$, and CC$\_160=1.010$. The SPIRE bands only trace the cold dust emissions and have  CC$\_250=0.94$, CC$\_350=0.92$, and CC$\_500=0.89$.

For each band, we also measure the random noise level from the emission-free region. The results are shown in Table A.1.. 

\begin{table*}[b]
{\scriptsize
\centering
\begin{minipage}{140mm}
\caption{The properties of the continuum data.}
\begin{tabular}{lcccccc}
\hline\hline
image data         &  rms noise            &  Absolute      &  resolution   &  pixel size   &  peak intensity$^a$  &  average intensity$^a$   \\
\quad              &  mJy pix$^{-1}$       &  flux error    &  (arcsec)     &  (arcsec)     &  mJy arcsec$^{-2}$   &  mJy arcsec$^{-2}$       \\
\hline                                                                                         
IRAC $8~\um$       &  0.2                  &  3\%           &  2.0          &  0.6          &  489                &  150   \\
MSX $12~\um$       &  0.5                  &  3\%           &  20           &  6            &  82                 &  40   \\
MSX $16.5~\um$     &  0.4                  &  5\%           &  20           &  6            &  87                 &  52   \\
MSX $21~\um$       &  3.0                  &  5\%           &  20           &  6            &  87                 &  52   \\
PACS $70~\um$      &  5                    &  10\%          &  9            &  2            &  $12\times10^3$      &  $5.3\times10^3$   \\
PACS $100~\um$     &  5                    &  10\%          &  6.7          &  1.6          &  $11\times10^3$      &  $2.0\times10^3$   \\
PACS $160~\um$     &  6                    &  10\%          &  11           &  3            &  $6\times10^3$       &  $1.5\times10^3$   \\
SPIRE $250~\um$    &  10                   &  5\%           &  18           &  6            &  $3\times10^3$       &  $0.8\times10^3$   \\
SPIRE $350~\um$    &  10                   &  5\%           &  25           &  10           &  $1\times10^3$       &  $0.3\times10^3$   \\
SPIRE $500~\um$    &  5                    &  5\%           &  36           &  14           &  $0.1\times10^3$     &  $0.2\times10^3$   \\
SCUBA $450~\um$    &  10                   &  30\%          &  9            &  2            &  810                &  400  \\
SCUBA $850~\um$    &  0.1                  &  10\%          &  14           &  1            &  50                 &  20   \\
\hline
\end{tabular} \\
$a.${ The intensity units are coverted into mJy arcsec$^{-2}$ in order to provide a fair comparison for all the bands. The images are kept in original appearance and not smoothed, re-gridded, or performed with any other computation. The peak and average intensities represented the values estimated within the observed region as shown in Figure 1.} \\
\end{minipage}
}
\end{table*}
 

\subsection{The SED fitting}
The SED fitting is based on the least square fit of the observed flux densities, that is to look for minimum of   
\begin{equation}
\chi^2=\sum_{\lambda} [\log S_\lambda(T_{\rm d})-\log S_{\lambda,{\rm obs}}]^2,
\end{equation}
$S_\lambda(T_{\rm d})$ is defined in Equation (1), and the free parameters to fit are $T_{\rm d}$, $\beta$, and $M_{\rm core}$. The fitting is implemented using the NonLinearLSQFitter in the Astropy package \footnote{\scriptsize Astropy is a community Python package for astronomy \citep{bray14}.}.
The errors in $\beta$, $T_{\rm d}$, and $M_{\rm core}$ are estimated by varying the data points within the uncertainty range for the fitting. a $\delta T_{\rm d}\simeq \pm 1$ K, $\delta \beta \simeq \pm 0.3$, and $50-60\%$ for $M_{\rm core}$. In the case that the SED contains two temperature components with comparable intensities, the wavelength range of $\lambda=160$ to 850 $\um$ is found to be still dominated by the cold component, while the $\lambda=20$ to 70 $\um$ is dominated by the hot component. In fact, the two components are only blended around FIR-5. The other cores are dominated by the cold dust while the emission out of the filament is dominated by the hot dust.


\subsection{The fitting of the \nht\ hyperfine structures}
In modelling the \nht\ spectra, we first calculate the \nht\ optical depth over the spectral frequency range

\begin{equation}
\tau(\nu)=\tau_0 \sum_{j=1}^{N} a_j \exp[-4\ln 2(\frac{\nu-\nu_0-\nu_j}{\Delta \nu})^2],
\end{equation}
wherein $\nu_j$ and $a_j$ are the rest frequency and relative intensity for each hyperfine component, respectively. $N$ is the total number of the hyperfine transitions, and $\Delta \nu$ is the FWHM line width. The values of $a_j$ and $N$ for (1,1) and (2,2) lines can be obtained from \citet{kukolich67}. $\tau_0$ is the integrated optical depth of all the components. For a given $(J,K)$ inversion transition, the modelled brightness temperature $T_{\rm b}$ is \citep{friesen09} 
\begin{equation}
T_{\rm b}(J,K)=\Phi \eta [J(T_{\rm ex})-J(T_{\rm bg})][1-\exp(-\tau_\nu)],
\end{equation}
where $\Phi$ is the beam filling factor for the emission region which is adopted as $\Phi=1$. $\eta$ is the mean-beam efficiency which is also adopted to be 1.0 if there is a flux calibrator. $T_{\rm ex}$ is the line excitation temperature. $T_{\rm bg}=2.73$ K is the temperature of the cosmic microwave background, $J(T)=(h\nu /k)[\exp(h\nu/kT)-1]^{-1}$ is the Planck-corrected brightness temperature. Frequency $\nu$ is related to the radial velocity $v$ as $(\nu-\nu_0)/\nu=(v-v_{\rm sys})/c$. The free parameters for the fitting are $T_{\rm ex}$, $\nu_0$, $\Delta \nu$, and $\tau_0$. The best-fit spectrum is obtained also using NonLinearLSQFitter.

The column density of the $(J,K)$ level is calculated following Equation (13) in \citet{rosolowsky08}:
\begin{equation}
\begin{aligned}
N(J,K)= & \frac{8\pi\nu_0^2}{c^2}\frac{g_1}{g_2}\frac{1}{A(J,K)} \\
        & \times \frac{1+\exp(-h\nu_0/kT_{\rm ex})}{1-\exp(-h\nu_0/kT_{\rm ex})}\int\tau(\nu)d\nu,
\end{aligned}
\end{equation}
wherein $A(J,K)$ is the Einstein A coefficient. Using the partition function Z,
\begin{equation}
Z=\sum_{J} (2J+1)S(J)\exp \frac{-h[BJ(J+1)+(C-B)J^2]}{k T_{\rm rot}},
\end{equation}
the total \nht\ column density is $N(1,1)\times Z/Z(1,1)$, wherein $B=298117$ MHz and $C=186726$ MHz are the rotational constants. The function $S(J)$ accounts for the extra statistical weight of the ortho- over para-\nht\ states, with $S(J)=2$ for $J=3$, 6, 9, ... and $S(J)=1$ for all other J values. 

The previous studies \citep[e.g.][]{tafalla04} presented an empirical formula for $T_{\rm k}$ as a function of $T_{\rm rot}$. But this relation might be inaccurate at high temperatures ($T_{\rm k}>20$ K) and densities. To accurately determine $T_{\rm k}$, we calculated the energy level distributions using the Non-LTE radiative transfer model RADEX \citep{vdtak07}. The input parameters includes $T_{\rm k}$, $n_{\rm tot}$, $N$(\nht) and $\Delta V$. We considered the $T_{\rm k}$ range from 15 to 30 K. The other parameters are set to be the values measured in Cd-2. With the input parameters, RADEX will estimate energy level population, and the theoretical $T_{\rm rot}$ is determined by the ratio of $N(2,2)/N(1,1)$ assuming a Boltzmann distribution:
\begin{equation}
\begin{aligned}
\frac{N(2,2)}{N(1,1)}= & \frac{g(2,2)}{g(1,1)}\exp\left(\frac{-\Delta E}{T_{\rm rot}}\right),
\end{aligned}
\end{equation}
The $T_{\rm rot}-T_{\rm k}$ relation can then be numerically sampled for a series of $T_{\rm k}$ values in our range. The result is shown in Figure A1. It shows that the derived relation for FIR-3 is very close to $T_{\rm rot}=T_{\rm k}$. This is within our expectation since at such high densities ($10^7$ \cc) the level population would be nearly thermalized.

\begin{figure}
\centering
\includegraphics[angle=0,width=0.7\textwidth]{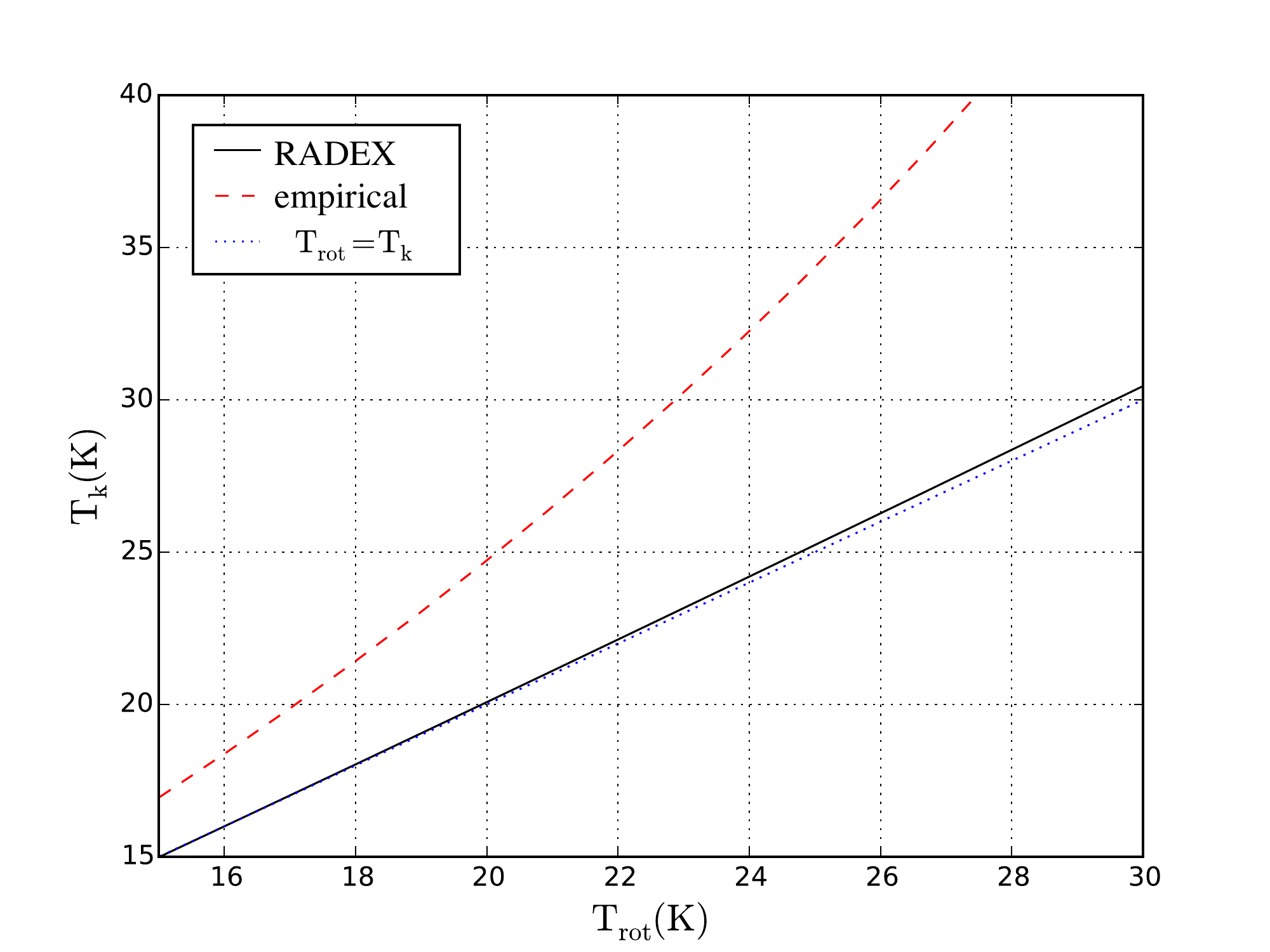} \\
\caption{\small The relation between the gas kinematic temperature $(T_{\rm k})$ and the rotational temperature $(T_{\rm rot})$ for the \nht\ (1,1) and (2,2) levels as numerically sampled by the RADEX program. The reference line of $T_{\rm rot}=T_{\rm k}$ is also plotted. }    
\end{figure}


\clearpage
\end{document}